\documentclass{ws-procs10x7}

\begin{document}

\title{Recent progress on Higgs physics at a $\gamma\gamma$ collider}

\author{Stephen Godfrey$^{\lowercase{a}}$ and Shouhua Zhu$^{\lowercase{a,b,c}}$}

\address{ $^a$Ottawa-Carleton Institute for Physics, 
Department of Physics, Carleton University, \\
Ottawa, Canada K1S 5B6\\
$^b$Institute of Theoretical Physics,
School of Physics, Peking University, \\
Beijing 100871, P.R. China\\
$^c$Institute of Theoretical Physics, Academia Sinica, Beijing
100080, P.R. China}

\twocolumn[\maketitle
\abstract{We review some recent developments in 
Higgs physics at a $\gamma\gamma$ collider.
We begin with single Higgs
boson production in $\gamma\gamma$ collisions proceeding via the
hadronic content of the photon. 
For SM Higgs masses of current theoretical interest,
the resolved photon contributions are non-negligible in precision
cross section measurements. We found that production of the
heavier Higgs bosons, $H^0$ and $A^0$, of the MSSM can probe regions
of the SUSY parameter space that will complement other
measurements. We showed that associated $t H^\pm$ production
in $\gamma\gamma$ collisions can be used to make an accurate
determination of $\tan\beta$ for low and high $\tan\beta$ by
precision measurements of the $\gamma\gamma \to H^\pm t +X $ cross
section. We then reviewed recent progress on Higgs physics in
direct photon processes in which the Higgs bosons are
produced via virtual loops. Precision measurements 
of the magnitude and phase of the $H\gamma\gamma$ coupling 
at a photon collider
can be used to determine Higgs parameters related to physics 
beyond the SM.}
]


\section{Categories of Higgs physics at a $\gamma\gamma$ collider}\label{subsec:prod}

The photon-photon ``Compton Collider'', which utilizes laser light
backscattered  off of highly energetic and possibly polarized
electron beams, has been advocated as a valuable part of the LC
physics program \cite{telnov}. In this talk we review some recent
progress on Higgs physics in: (1) resolved photon processes and
(2) direct photon processes. Due to space limitations, some
important issues cannot be included, for example, Higgs pair
production which is crucial for the measurement of the triple Higgs
coupling.

We begin by exploring various aspects of Higgs boson production
in resolved photon processes, i.e. via the hadronic content of the
photon,\cite{Doncheski:2001uh,Doncheski:2003te} which we briefly
describe.  Details and complete references are given in Ref. 2 and
3. In the resolved photon approach the quark and gluon content of
the photon are treated as partons described by partonic
distributions, $f_{q/\gamma}(x,Q^2)$, in direct analogy to partons
inside hadrons\cite{fph}. The parton subprocess cross sections
are convoluted with the parton distributions to obtain the final
cross sections. These have to be further convoluted with the
photon energy distributions obtained from either the backscattered
laser or the Weizs\"{a}cker Williams distributions to obtain cross
sections that can be compared to experiment.

We then review recent progress on Higgs physics in direct
photon processes.\cite{HF1,HF2,Niezurawski:2002jx,Niezurawski:2004ui,Godbole:2004xe,NZK739}
It is well known that one of the strongest motivations for the
Compton collider is to measure Higgs boson properties via single
neutral Higgs production through loops. Measurement of the
$\gamma\gamma \to H$ cross section
is especially interesting as it proceeds via loop contributions
and is therefore sensitive to new particles that cannot be
produced directly. Based on the precise measurements of the
magnitude and phase of $H\gamma\gamma$ at a photon collider, Higgs 
parameters for new physics beyond the SM can be precisely determined. 

\section{Resolved Photon processes}\label{subsec:wpp}

\subsection{Single neutral Higgs production}

Due to the importance of single neutral Higgs boson production via
$\gamma\gamma \rightarrow H$, it is important that all SM
contributions to this process be carefully considered. The
resolved photon contributions to Higgs production are shown in
Fig. 1 for the backscattered laser case with
$\sqrt{s_{ee}}=500$~GeV. We also show the $\gamma\gamma\to
H$ cross section which proceeds via loops and the contribution from
gluon fusion, $\hat{\sigma}(gg\to H)$, which arises from the gluon
content of the photon.  A final process  is $\gamma\gamma\to H
W^+W^-$ whose cross section is comparable to the resolved photon
processes. Although the loop process dominates over the resolved
photon processes for the full range of Higgs masses,
the 
latter processes contribute at the percent level for $M_H\sim
150$~GeV and $\sqrt{s_{ee}}=500$~GeV increasing to several percent
for $\sqrt{s_{ee}}=1.5$~TeV.
Thus, 
Higgs production via the hadronic content of the photon may not be
negligible for precision measurement of $\sigma(\gamma\gamma \to
H)$ suggesting that these contributions deserve further study.

\begin{figure}
\begin{center}
\centerline{\epsfig{file=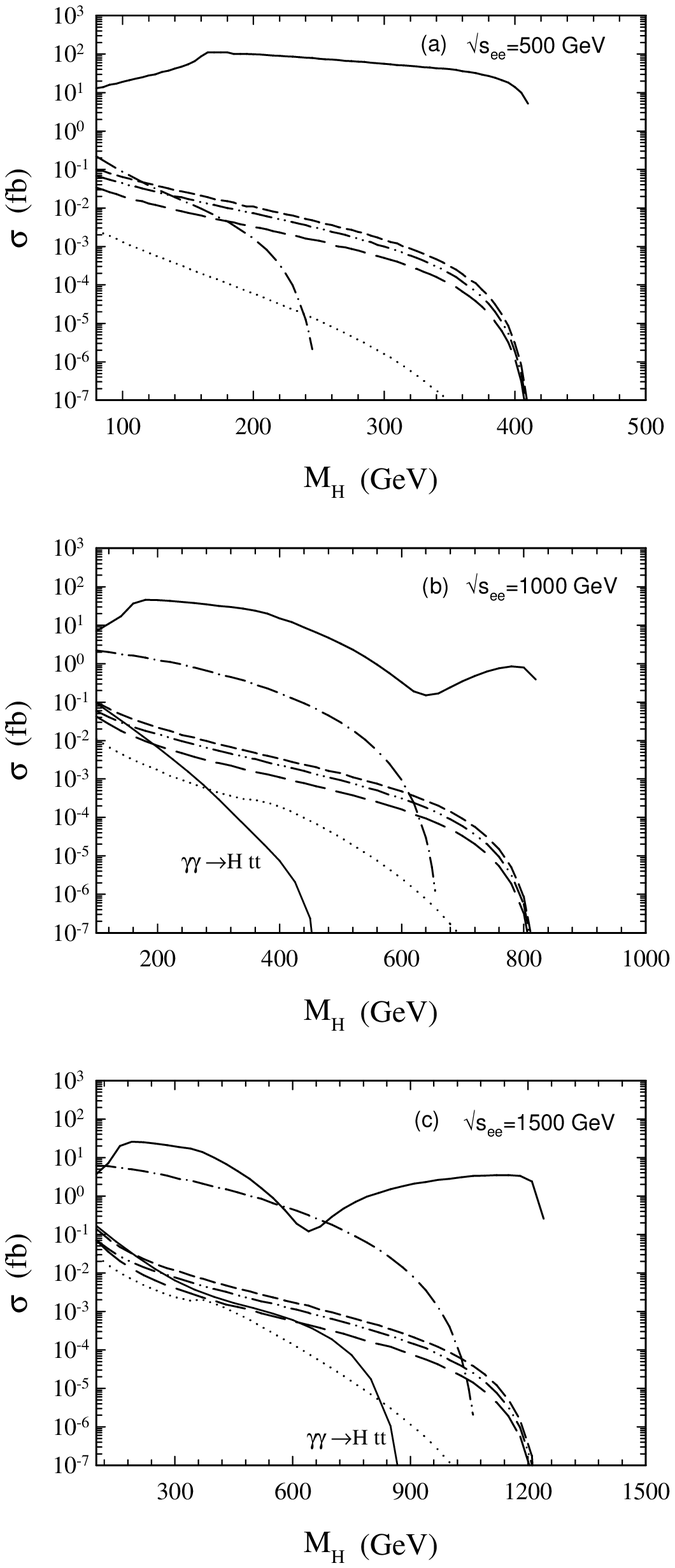,width=7.4cm,clip=}}
\caption{Production cross sections for SM Higgs boson including
the backscattered laser spectrum.  The solid line is for
$\gamma\gamma\to h$, the short dashed line for
$\hat{\sigma}(b\bar{b}\to h) +\hat{\sigma}(c\bar{c}\to h)$, the
dot-dot-dashed line for $\hat{\sigma}(c\bar{c}\to h)$, the
long-dashed line for $\hat{\sigma}(b\bar{b}\to h)$, the dotted
line for $\hat{\sigma}(gg\to h)$ and the dot-dashed line for
$\hat{\sigma}(WW\to H)$.} 
\end{center}
\end{figure}


In the MSSM there exist a total of three neutral Higgs bosons
which can be produced in $\gamma\gamma$ collisions; $\gamma \gamma
\to h^0, \; H^0, \; A^0$. However, the resolved photon process has
different dependence on $\tan\beta$ than that of the loop
processes. If $H$ and $A$ were produced in sufficient quantity the
cross section could be used to constrain $\tan\beta$. This can be
seen most clearly in Fig. 2 which shows the regions of the
$\tan\beta - M_{A}$ plot which can be explored via $A$ production
for $\sqrt{s}_{ee}=500$~GeV.  The regions covered would complement
measurements made in other processes.

\begin{figure}
\begin{center}
\centerline{\epsfig{file=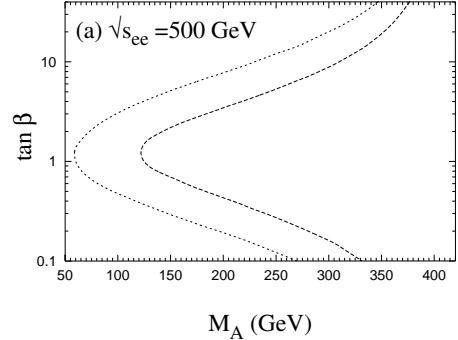,height=6.0cm,angle=-90}}
\caption{ Regions of sensitivity in $\tan\beta -M_A$ parameter
space to $A$ production via resolved photons with backscattered
laser photons for $\sqrt{s_{ee}}=500$~GeV. The region to the left
is accessable and the region to the right is unaccessable. The
dotted line gives  $\sigma = 0.1$~fb contour so that at least
100~events would be produced in the region to the left for
1~ab$^{-1}$ integrated luminosity. Similary the dashed line gives
the $\sigma = 0.02$~fb contour  designating the the boundary for
producing at least $20$~events. }
\end{center}
\end{figure}

\subsection{Measurement of  $\tan\beta$ in associated $tH^\pm$ production}
The ratio of neutral Higgs field vacuum expectation values,
$\tan\beta$, is a key parameter needed to be determined in type-II
Two-Higgs Doublet Models and the MSSM. $\tan\beta$ can be measured
in associated $tH^\pm$ production in $\gamma\gamma$ collisions.
The subprocess $b\gamma \to H^- t$ utilizes the $b$-quark content
of the photon. $\tan\beta$ enters through the $tbH^\pm$ vertex.
In Fig. 3 we show the cross section as a function of $\tan\beta$
with the measurement precision superimposed. We found that
$\gamma\gamma\to tH^\pm +X$ can be used to make a good
determination of $\tan\beta$ for most of the parameter space with
the exception of the region around $\tan\beta\simeq 7$ where the
cross section is at a point of inflection.  This measurement
provides an additional constraint on $\tan\beta$ which complements
other processes.

\begin{figure}
\begin{center}
\centerline{\epsfig{file=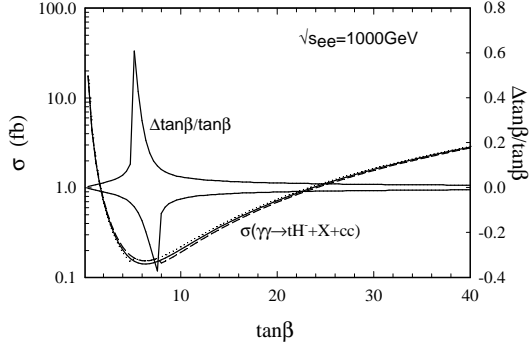,width=7.0cm,clip=}}
\caption{ $\sigma(\gamma\gamma\to t H^- +X)$ {\it vs.} $\tan\beta$
for the backscattered laser case and the sensitivities to
$\tan\beta$ based only on statistical errors (solid lines) for
$\sqrt{s}_{ee}=1$~TeV and $M_H = 200$~GeV.  For the cross
sections, the solid line represents the expected cross section at
the nominal value of $\tan\beta$, while the dashed (dotted) line
represents the expected cross section at $\tan\beta -
\Delta\tan\beta$ ($\tan\beta + \Delta\tan\beta$).}
\end{center}
\end{figure}

\subsection{Other resolved photon process}

 Choi {\it et al} \cite{Choi:2004ne} studied the
related process of $\tau\tau \to H$ where the $\tau$'s come from
the fermionic content of the photon. They show that $\Delta
\tan\beta \sim 0.9 \ to\ 1.3$ is uniform in $\tan\beta$ for all $M_A$
up to the kinematic limit, and the results are encouraging enough to
start real experimental simulations.

\section{Direct photon processes}

As discussed above, the neutral Higgs boson can be produced via
loops which is sensitive to new virtual particles arising in new
physics beyond the SM. This process can be used to measure the
$H\gamma\gamma$ magnitude from $\Gamma_{\gamma\gamma}$ and the phase
$\Phi_{\gamma\gamma}$. According to the decay products of the
Higgs boson, the investigations can be classified into: (1)
$H\rightarrow \ heavy \ fermions$ and (2) $H\rightarrow VV$ with
$V=W$ and $Z$.

\subsection{ $H\rightarrow \ heavy \
fermions$ }

Usually the couplings of the Higgs boson with fermions are
proportional to the fermion's mass, which is what makes the heavy fermion
decay modes interesting\cite{HF1,HF2}. For example, for $H
\rightarrow b \bar b$, a precision of 2-9\% can be achieved on
cross section measurements after one year of Photon Collider (PC) 
running for a SM
Higgs mass between 120 to 160 GeV, while for MSSM Higgs
bosons a measurement precision of 11-23\% on the cross 
sections can be achieved for $M_A=200\sim 350$ GeV.\cite{HF1}

\subsection{$H\rightarrow VV$ with $V=W$ and $Z$}

In the SM,  from simultaneous fits to the $WW$ and $ZZ$ mass
spectra, $\Gamma_{\gamma\gamma}$ and $\Phi_{\gamma\gamma}$
can be measured with precision of $\sim$ 4-9\% and 40-120 mrad
respectively\cite{Niezurawski:2002jx}.

In the SM-like two Higgs doublet model [2HDM(II)] there is
only one parameter, $\tan\beta$. But one can also
introduce the additional CP-violating $H-A$ mixing parameter $\Phi_{HA}$.
Both $\tan\beta$ and $\Phi_{HA}$ can be measured to $\sim$ 10\%
(for $\Phi_{HA} \sim 0$) and 0.1 rad if $\tan\beta$ is not too 
large\cite{Niezurawski:2004ui}.

In the general 2HDM (II) one defines
$$
\chi_x\equiv \frac{g_{Hxx}}{g_{Hxx}^{SM}}, {\rm with} \ H=H,h^0,A.
$$
LHC measurements can constrain $\chi_U$ (Higgs
boson production  via top virtual loop where $U$ is for up-type quarks), 
the LC can constrain $\chi_V$
(Higgs boson radiating off $Z$ boson) and the PC can constrain both
(Higgs boson production via gauge boson and quark loops). The
combined analysis from the LHC, LC and PC, is depicted in Fig. 4. The
average errors are:
\begin{eqnarray}
<\Delta \chi_V>=0.033 \\
<\Delta \chi_U>=0.12\\
<\Delta \Phi_{HA}>=150 \ mrad
\end{eqnarray}
for $\chi_V=0.7$, $\chi_U=-1$ and $m_H=250$
GeV\cite{Godbole:2004xe}.

\begin{figure}
\begin{center}
\centerline{\epsfig{file=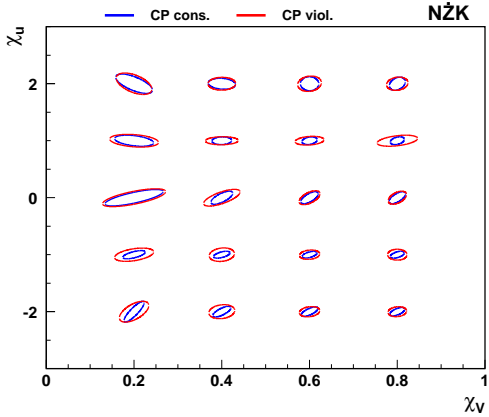,width=7.0cm,clip=}}
\caption{ Taken from Ref. 9. 
Expected total
error contours (1$\sigma$) in the determination of the basic
Higgs-boson couplings to vector bosons ($\chi_V$) and up fermions
($ \chi_u $) from combined fit to invariant mass distributions
measured at LHC, LC and Photon Collider, for the CP-conserving
(blue) and CP-violating (red) 2HDM~(II). Production and decays of
heavy Higgs boson are considered for $M_H=250$ GeV. }
\end{center}
\end{figure}

In the generic model the $HVV$ couplings, $g_{HZZ}$ and $g_{HWW}$,
can be written as
\begin{eqnarray}
&& ig \frac{m_Z}{\cos\theta_W} \left( g_{\mu\nu} \lambda_H+
\lambda_A \epsilon_{\mu\nu\rho\sigma}
\frac{P^{\rho\sigma}}{m_Z^2} \right), \\
&& ig m_W \left( g_{\mu\nu} \lambda_H+ \lambda_A
\epsilon_{\mu\nu\rho\sigma} \frac{P^{\rho\sigma}}{m_W^2} \right)
\end{eqnarray} respectively, with $P^{\rho\sigma}=(p_1+p_2)^\rho
(p_1-p_2)^\sigma$, $\lambda_H=\lambda \cos\Phi_{CP}$ and
$\lambda_A=\lambda \sin\Phi_{CP}$. In the SM, $\lambda_H=1$ and
$\lambda_A=0$. Detailed simulations, especially on the angular
distributions, show that $\Delta \Phi_{CP}=50\ mrad$ and $\lambda$
can be measured to several percents \cite{NZK739}.


\section{Summary}
We reviewed recent progress on Higgs physics
at a photon colliders, especially Higgs physics in resolved
photon processes. We can see that there is rich physics at the photon
collider and much more investigation needs to be done.

\section*{Acknowledgments}
This work was supported in part by the
Natural Sciences and Engineering Research Council of Canada. We
thank P.~Niezurawski, A.~F.~Zarnecki and M.~Krawczyk for
contributions to this talk and P. Kalyniak and R. Hemingway for
helpful comments.

\end{document}